\documentclass{epl}
\usepackage{graphicx}
\usepackage{amsmath, amsfonts, amssymb, bm}

\begin{document}

\title{Double-EIT ground-state laser cooling without blue-sideband heating}
\shorttitle{Double-EIT ground-state laser cooling...}
\author{J\"org Evers\inst{1} \thanks{evers@physik.uni-freiburg.de}
 and Christoph H. Keitel\inst{2,1} \thanks{keitel@mpi-hd.mpg.de}}
\institute{
\inst{1}Theoretische Quantendynamik, Fakult\"at f\"ur Physik und Mathematik,
Universit\"at Freiburg, Hermann-Herder-Stra{\ss}e 3, 79104 Freiburg, Germany\\
\inst{2}Max-Planck-Institut f\"ur Kernphysik, Saupfercheckweg 1, 69117
Heidelberg, Germany}
\pacs{pacs}{32.80.Pj, 42.50.Gy, 42.50.Vk}

\date{\today}
\maketitle

\begin{abstract}
We discuss a laser cooling scheme for trapped atoms or ions
which is based on double electromagnetically induced transparency (EIT) 
and makes use of a four-level atom in tripod configuration.
The additional fourth atomic state is coupled by a strong coupling 
laser field to the usual three-level setup of 
single-EIT cooling. This effectively allows to
create two EIT structures in the absorption spectrum of the system to be cooled,
which may be controlled by the coupling laser field parameters 
to cancel both the carrier- and the blue-sideband excitations.
In leading order of the Lamb-Dicke expansion, this 
suppresses all heating processes. As a consequence, the double-EIT 
scheme can be used to lower the cooling limit by almost two powers of the 
Lamb-Dicke parameter as compared to single-EIT cooling. 
\end{abstract}

\section{\label{sec-intro}Introduction}
Many current experiments involving
the preparation or manipulation of
atoms and ions require a precise
coherent control of the system of interest.
This does not only apply to internal 
degrees of freedom, but also to the external
motional degrees of freedom. In the last few
years, the laser cooling of trapped ions
or atoms has therefore been a subject of
intense research and is now a routine
tool in many laboratories.
Starting from the first observation
of laser cooling~\cite{erste},
many interesting applications have been
made possible by laser cooling. Examples
are the direct observation of quantum
jumps~\cite{jumps}, the preparation of
atoms in the motional ground state~\cite{ground-state},
and high-precision spectroscopy~\cite{spectroscopy}.

Apart from cooling on dipole-forbidden transitions~\cite{ground-state,forbidden}
and cooling by stimulated Raman transitions~\cite{raman-cooling}, cooling
facilitated by electromagnetically induced transparency (EIT)~\cite{eit}
is a promising recent technique to reach the mechanical
ground state. EIT cooling has already been observed 
experimentally~\cite{eit-exp}.
Once a trapped atom reaches the Lamb-Dicke regime
e.g. by unresolved Doppler cooling,
the interaction with a cooling laser field
may excite the atom by three kinds of processes.
Carrier excitations involve an excitation
of the internal electronic degree of freedom 
of the atom without a change of the motional quantum number.
Processes where simultaneously with the internal excitation the 
motional quantum number is increased (decreased) are known as
blue (red) sideband excitations.
Both excitation-deexcitation cycles involving 
carrier- and blue-sideband excitations
may lead to a heating of the trapped system.
Red-sideband excitations may cool the system.
The basic idea of EIT cooling is to design the
absorption properties of the sample to be cooled
such that carrier excitations are inhibited,
and such that red-sideband transitions dominate
over blue-sideband transitions.
The cooling limit is then reached
if the heating and the cooling rates become
equal, and is typically above zero.
\begin{figure}[b]
\center
\includegraphics[height=3cm]{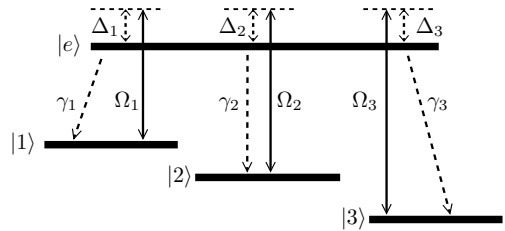}
\caption{\label{fig-system}Level scheme for the
double-EIT cooling scheme. The upper state $|e\rangle$
is connected to three lower states $|i\rangle$
($i\in \{1,2,3\}$) by dipole-allowed transitions.
$\Omega_i$ and $\Delta_i$ are the respective
Rabi frequencies and detunings of the driving laser fields,
and  $\gamma_i$ are spontaneous decay rates.}
\end{figure}
However the incoherent thermal 
motion is a source of decoherence and disturbs the precise 
coherent control of the system.
One way to resolve this problem is to find control
schemes for ``hot'' systems which do not require
a cooling to the ground state~\cite{hot}.
Nevertheless many recent theoretical proposals rely
on an atomic or ionic system cooled to the
motional ground state in order to allow for a 
precise control and to avoid decoherence.
Typical examples for this are are quantum communication 
schemes or  quantum information 
processing.

Therefore in this article we propose a
``double-EIT'' cooling scheme which 
in addition to the three-level $\Lambda$-setup in the usual EIT scheme
makes use of an additional third ground state. This state is coupled 
by a strong coupling laser to the upper state, see Fig.~\ref{fig-system}.
This effectively allows to create two independent EIT-structures
in the absorption spectrum of the trapped atom, which can
be controlled by the two coupling laser fields to cancel
the carrier- and the blue-sideband transitions, respectively.
With this technique, the dominant heating processes of
single-EIT cooling are suppressed in double-EIT cooling.
Thus  the predicted cooling limit of the double-EIT
scheme is suppressed by a factor of order $\eta^2$
as compared to conventional EIT cooling, where $\eta$ is the
Lamb-Dicke parameter. 
For many cooling setups, double-EIT cooling 
does not require more
effort than single-EIT, as the upper state decays to more than
two lower states as needed for single-EIT cooling. Instead of 
repumping the system arbitrarily, a suitable choice of 
the pump field leads to double-EIT cooling and thus to improved cooling
performance. Therefore with the double-EIT scheme the unwanted additional 
decay pathways become an advantage.

\section{\label{sec-system}The double-EIT scheme}
The system consists of a trapped atom confined in a harmonic 
potential with trap frequency $\nu$ and mass $M$. 
Internally, the atom has one upper state $|e\rangle$ and three
lower states $|i\rangle$ ($i\in \{1,2,3\}$) which are connected
by dipole-allowed transitions to the upper state (see Fig. \ref{fig-system}). Each of the
transitions is driven by a laser field with Rabi frequency
$\Omega_i$ and detuning $\Delta_i = \omega^L_i - (\omega_e - \omega_i)$ 
($i\in \{1,2,3\}$) where $\omega^L_i$ is the frequency of the $i$th
laser field and $\hbar \omega_j$ ($j\in \{e,1,2,3\}$) is the energy of 
the atomic state $|j\rangle$. The upper state may further decay
to the lower states with the decay rates $\gamma_i$, respectively.
Thus the system Hamiltonian for the coherent evolution is given by~\cite{rmp,review2}
\begin{eqnarray}
H&=&\hbar \nu a^{\dagger} a + \hbar \sum_{j=1}^{3} \Delta_j\:|j\rangle\langle j|
+\hbar \sum_{j=1}^{3}\frac{\Omega_j}{2}\left ( e^{-ik_j\cos(\phi_j)x} |e\rangle\langle j| + h.c.\right )\, ,
\end{eqnarray}
where $k_j$ ($j \in \{1,2,3\}$) are the wave vectors of the coupling laser
field traveling in directions given by the angles $\phi_j$, $x$ is the 
position operator, and $a$($a^\dagger$) is the annihilation (creation) 
operator for a motional quantum.
The master equation for the system density operator $\rho$ reads
$\frac{\partial}{\partial t} \rho = \frac{1}{i\hbar} 
\left [ H, \rho \right ] + {\mathcal L}_{\rm SE}\rho$.
Here, the super-operator ${\mathcal L}_{\rm SE}$ describing the spontaneous decays with angular distribution
$\mathcal{N}_j(\theta)$ ($j=1,2,3$) acts as
($[.,.]_+$ denotes the  anti-commutator)~\cite{rmp}
\begin{figure}[b]
\center
\includegraphics[height=4cm]{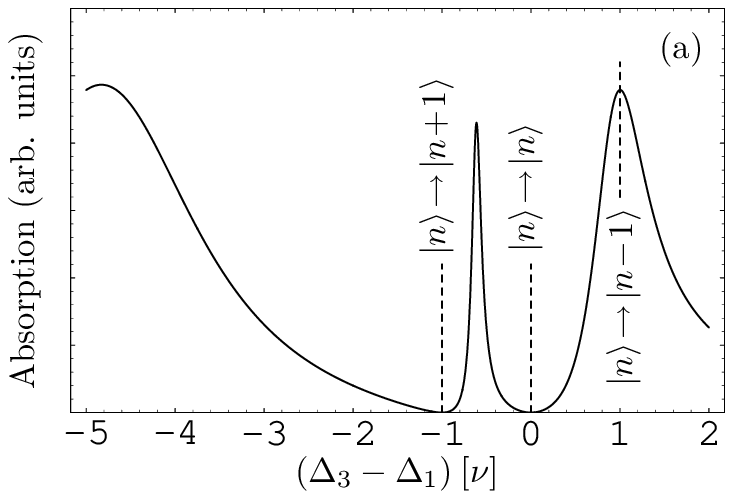}
\hspace*{1cm}
\includegraphics[height=4cm]{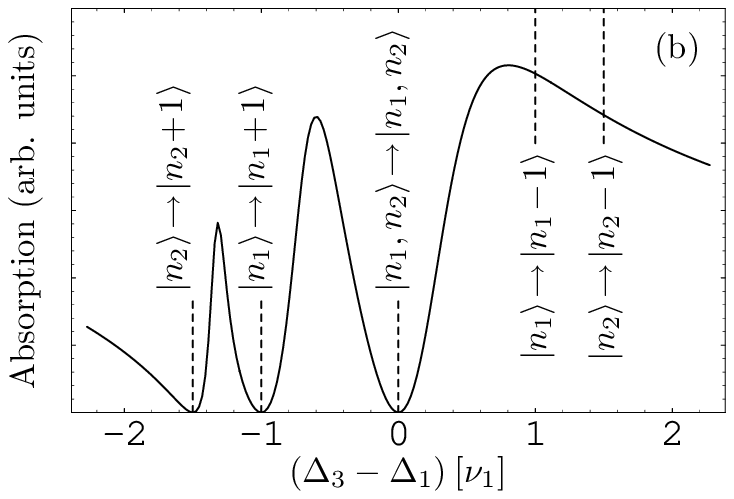}
\caption{\label{fig-absorb}Absorption rate of the cooling laser. 
(a) Double-EIT. As there are two EIT structures, the absorption vanishes 
at two values for the
detunings $\Delta_3$ of the cooling laser.
The vertical lines
mark the positions of these zeros and of the maximum of the
absorption. For appropriately chosen parameters, the carrier transition
and the blue-sideband heating are completely canceled, while the
red-sideband cooling is maximized. Here, the parameters are
$\Omega_1 = \Omega_2 = \gamma_3, \Omega_3 = \gamma_3/20, 
\Delta_1 = \gamma_3$. The other parameters are chosen to satisfy
Eq.~(\ref{c1}-\ref{c3}). The trap frequency is $\nu \approx 0.3 \gamma_3$.
(b) Absorption spectrum for simultaneous cooling at two different trap frequencies 
($\nu_2 = 1.5\:\nu_1$) with triple-EIT.}
\end{figure}
\begin{equation}
{\mathcal L}_{\rm SE}\rho = 
-\sum_{j=1}^{3} \frac{\gamma_j}{2} \left ( \bigl 
[|e\rangle\langle e| ,\rho \bigr ]_+
- 2 \int_{-1}^{1}  \mathcal{N}_j(\theta) \: |j\rangle\langle e| 
e^{ik_j\cos(\theta) x} \rho e^{-ik_j\cos(\theta) x} |e\rangle\langle j| 
\:  \textrm{d}\!\cos(\theta) \right ) \, .
\end{equation}
In the first step of the analysis we neglect the external degrees 
of freedom~\cite{eit}.
Throughout the analytical calculations, we only consider the spontaneous decay on the 
cooling transition $3\leftrightarrow e$ with the rate $\gamma_3$. 
The other decay rates $\gamma_1, \gamma_2$ are taken into account in the 
numerical analysis. 
The internal state dynamics of the atoms is then given by the
following equations of motion for the density matrix $\rho$~\cite{n-lower-state}:
\begin{eqnarray}
\dot{\rho}_{jj} &=& i \frac{\Omega_j}{2} \left (\rho_{je}-\rho_{ej} \right) 
   + \delta_{j3} \gamma_3 \rho_{ee} \:,\\
\dot{\rho}_{je} &=& -\left (i \Delta_j + \delta_{j3}\frac{\gamma_3}{2}\right )\rho_{je} 
   - i \frac{\Omega_j}{2}\rho_{ee} 
   + i \sum_{l=1}^3 \frac{\Omega_l}{2}\rho_{jl} \:,\\
\dot{\rho}_{jk} &=& i (\Delta_k-\Delta_j)\rho_{jk} 
   - i \frac{\Omega_j}{2}\rho_{ek} 
   + i \frac{\Omega_k}{2}\rho_{je}\:,
\end{eqnarray}
for $j,k\in \{1,2,3\}$ with $j\neq k$. $\delta_{jk}$ is the Kronecker
delta, and the equations for the other density matrix elements are obtained
using the relations $\textrm{Tr}(\rho) = 1$ and $\rho_{jk}=\rho_{kj}^*$. 
In the following, the two 
transitions $1\leftrightarrow e$ and $2\leftrightarrow e$ 
are assumed to be driven by coupling laser fields which effectively 
dress the atom in order to modify the interaction 
with the third laser field on the cooling transition $3\leftrightarrow e$.
The cooling effect depends 
on the scattering rate of the cooling laser field.
A typical scattering spectrum of the cooling laser is shown in Fig.~\ref{fig-absorb}(a).
The spectrum may be understood as consisting of two Fano-like structures 
characteristic for EIT which may be controlled independently by the two strong 
coupling laser fields. For a three-level $\Lambda$-type atom, the 
corresponding scattering  spectrum only contains one such structure.

Taking into account the harmonic motion of the atom with 
frequency $\nu$,
the carrier excitation ($|3, n\rangle \to |e, n\rangle$) occurs at
$\Delta_3 = \Delta_1$. In the following, we assume these
detunings to be positive. The blue-sideband excitation 
due to the process $|3, n\rangle \to |e, n+1\rangle$
is at detuning $\Delta_3 = \Delta_1 - \nu$, as the
remaining energy is required to create 
a quantum of the harmonic motion. 
The red-sideband excitation ($|3, n\rangle \to |e, n-1\rangle$)
is at $\Delta_3 = \Delta_1 + \nu$ (see Fig.~\ref{fig-absorb}).
Thus it is clear that in order to achieve efficient cooling
of the system, the scattering spectrum has to be modified
such that it has minima at $\Delta_3 - \Delta_1 = 0, -\nu$
and a maximum at $\Delta_3 - \Delta_1 = \nu$.
In order to obtain conditions for the laser field parameters, 
we evaluate the scattering spectrum. We keep the cooling 
field to all orders,
as in general the cooling field is not assumed to be weak 
as compared to the coupling fields.
In the resulting expression, we set $\Delta_3 = \Delta_1 + \nu$
and $\Delta_2 = \Delta_1 - \nu$ and maximize with respect to
$\nu$. One of the solutions is the condition for $\nu$, as
the laser parameters need to be chosen such that the
scattering rate has its maximum value at $\Delta_3 = \Delta_1+\nu$.
This yields the following conditions for the 
laser parameters:
\begin{eqnarray}
\Delta_3 &=& \Delta_1 \, ,\qquad \Delta_2 = \Delta_1 - \nu\, ,  \label{c1} \\
\nu &=& \frac{1}{2}\left (\sqrt{\Delta_1^2+ \Omega_1^2 + \Omega_2^2/2 +\Omega_3^2} - \Delta_1  \right )\, . \label{c3}
\end{eqnarray}
In fact, these conditions have been used in Fig.~\ref{fig-absorb}(a).
For $\Omega_2=0$, the conditions
Eqs.~(\ref{c1},\ref{c3}) for the detuning and for $\nu$ reduce to
the corresponding result for the three-level case~\cite{eit2}.
The interpretation of these results may be given
in terms of the dressed state of the four-level 
atomic system driven by the two strong coupling
laser fields with Rabi frequencies $\Omega_1, \Omega_2$.
The maxima in the scattering spectrum shown in Fig.~\ref{fig-absorb}
correspond to the position of these dressed states, and the
widths of the peaks in the spectrum correspond to the 
dressed decay rates. The position of one of the narrow
dressed states has to be adjusted by the laser parameters
such that it is at $\Delta_3 + \nu$ and coincides with
the red-sideband transition frequency. The minima in the 
scattering spectrum
arise from two-photon resonance conditions of the
cooling laser field with each of the two coupling
laser fields at $\Delta_3 = \Delta_1$ and $\Delta_3 = \Delta_2= \Delta_1 - \nu$.

At second order perturbation theory in the Lamb-Dicke parameter,
the system evolution may
be described by a rate equation for the population
$\Pi (n)$ of the motional number states $|n\rangle$ 
which is given by~\cite{eit,rmp,review2}
\begin{eqnarray}
\frac{d}{dt} \Pi (n) =   A_- [ (n+1) \Pi (n+1) - n\Pi (n) ] 
     + A_+ [ n \Pi (n-1) - (n+1)\Pi (n) ] \, .
\end{eqnarray}
From this, one may obtain an equation for the time dependence of the  mean number of 
vibration excitations $\langle n \rangle = \sum_{k=0}^\infty \Pi(k)k$ which is given by
$\langle \dot{n} \rangle = - \left ( A_- - A_+ \right )\langle n \rangle + A_+ $,
with the cooling rate $W= A_- - A_+$
and the steady state value
$\langle n \rangle_{ss} = A_+/W$.
The coefficients $A_\pm$ can be obtained by expanding
the Hamiltonian describing the coherent interaction 
of the atom  with the laser fields with respect to the Lamb-Dicke parameter.
The first-order term of this expansion is
$V_1=\frac{i\hbar}{2} \sum_{j=1}^3\: k_j \cos (\phi_j) \Omega_j 
\left ( |j\rangle\langle e| - |e\rangle\langle j|\right )$.
The fluctuation spectrum of this operator is given by
\begin{equation}
S(\omega) = \frac{1}{2M\nu\hbar} \int_0^{\infty} \: e^{i\omega \tau} \: \langle V_1(\tau)V_1(0) \rangle_{ss} \: d\tau \: ,
\end{equation}
where the subindex ''ss'' denotes the steady state of the expectation value.
The coefficients $A_\pm$ are then
$A_\pm = 2 \textrm{Re} \left \{ S(\mp \nu) \right \} $ .
%
As in the single-EIT scheme with three-level $\Lambda$-type atoms, we set 
$\Delta := \Delta_1 = \Delta_3$ in order to fulfill Eq.~(\ref{c1}) and thus
to eliminate the carrier excitations. This yields
\begin{eqnarray}
\frac{A_\pm}{\eta^2} &=&  \frac{\Omega_1^2}{\Omega_1^2+\Omega_3^2}\:\frac{\gamma_3\nu^2\Omega_3^2}
{4 \left \{ [(\Omega_1^2+\Omega_3^2)/4 - \nu(\nu \mp \Delta)] + \mathcal{E_\pm} \right \}^2 + \gamma_3^2\nu^2},
\label{rates}
\end{eqnarray}
with
\begin{eqnarray}
\mathcal{E_\pm} = \mp \frac{\nu \Omega_2^2}{4(\Delta -\Delta_2 \mp \nu)} \, .
\end{eqnarray}
$\eta = \eta_1\cos(\phi_1)  - \eta_3 \cos(\phi_3) $ is the relevant Lamb-Dicke
parameter, where $\eta_i = k_i\sqrt{\hbar / 2M\nu}$ ($i=1,2,3$) are
the Lamb-Dicke parameters of laser field mode $i$.
This expression is the same as the 
corresponding result for
the single-EIT three-level case~\cite{eit2} 
except for the additional
contribution $\mathcal{E}_\pm$ in the denominator,
which depends on the parameters $\Omega_2, \Delta_2$ of the second
coupling laser field.
For $\Omega_2 = 0$, one has $\mathcal{E}_\pm=0$ and thus
obtains the result for the three-level case.
\begin{figure}[t]
\center
\includegraphics[height=4cm]{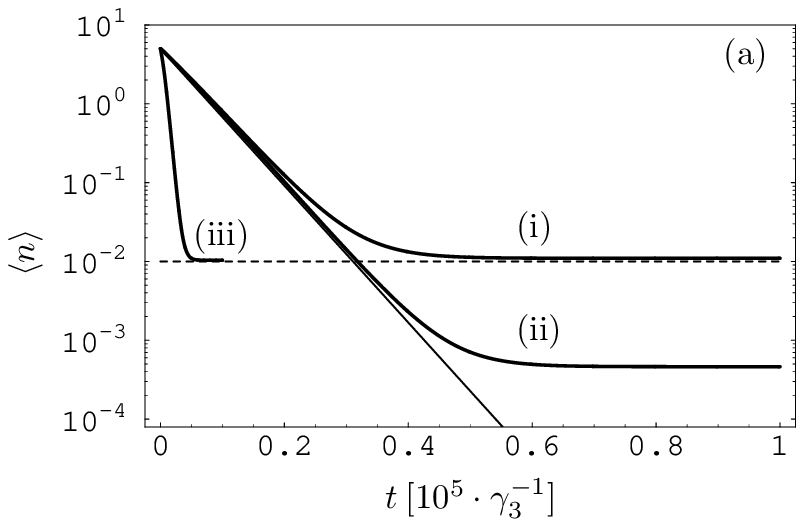}
\hspace*{0.5cm}
\includegraphics[height=3.9cm]{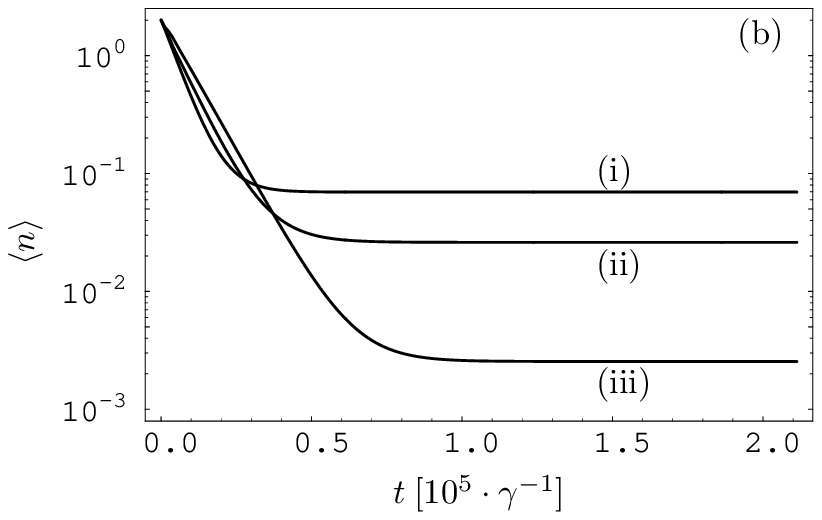}
\caption{\label{fig-mc}Numerical simulations of the cooling
dynamics. (a) Cooling of a Ca$^+$-ion. The parameters are $\nu=0.1\:\gamma_3$, $\Delta_1=\Delta_3=2.5\:\gamma_3$,
$\Delta_2=2.4\:\gamma_3$ and $\eta=0.145$. (i) Single-EIT cooling
with $\Omega_1=\gamma_3$ and $\Omega_3=0.1\:\gamma_3$. (ii) Double-EIT cooling
with $\Omega_1=0.8\:\gamma_3$, $\Omega_2=0.8944\:\gamma_3$ and $\Omega_3=0.1\:\gamma_3$. 
(iii) Double-EIT cooling with $\Omega_1 = \Omega_2 = \Omega_3 = 0.645\: \gamma_3$.
The dashed horizontal line marks the single-EIT cooling limit,
and the diagonal line shows an exponential decay with the single-EIT cooling rate for parameters as in (i). (b) EIT-cooling of Hg$^+$-ions. The parameters
are $\gamma = 69$~MHz, $\gamma_1 = \gamma_2 = \gamma_3 = \gamma/3$, 
$\nu= 1.5$~MHz,  $\Omega_3= 4$~MHz, $\Delta_1 = \Delta_3 = 80$~MHz, 
$\eta_1 \cos(\phi_1) = \eta_2 \cos(\phi_2)= 0.13$, 
$\eta_3 \cos(\phi_3)= -0.13$. (i) Single-EIT cooling with
repump field: $\Delta_2 = 0$, $\Omega_1 = 21$~MHz,
$\Omega_2=8$~MHz. 
(ii) Double-EIT cooling as in (i), but with $\Delta_2 = \Delta_1 - \nu$. 
(iii) Double-EIT cooling with $\Delta_2 = \Delta_1 - \nu$,
$\Omega_1 = 4$~MHz, $\Omega_2\approx 30$~MHz according to Eq.~(\ref{c3}).}
\end{figure}
On applying the other conditions on the laser
field parameters Eq.~(\ref{c1},\ref{c3}) for the optimum cooling conditions, 
the rates $A_\pm$ become
\begin{eqnarray}
A_+^c = 0 \, , \qquad 
A_-^c = \eta^2 \frac{\Omega_1^2}{\Omega_1^2+\Omega_3^2}\:\frac{\Omega_3^2}{\gamma_3} \, ,
\end{eqnarray}
such that the cooling rate and the cooling limit simplify to
\begin{eqnarray}
W^c = \eta^2 \frac{\Omega_1^2}{\Omega_1^2+\Omega_3^2}\:\frac{\Omega_3^2}{\gamma_3} \, , \qquad
\langle n \rangle_{ss}^c = 0 \, .
\end{eqnarray}
Here, the index ''c'' denotes the values obtained after
applying the conditions Eq.~(\ref{c1},\ref{c3}).
Thus in this order of the expansion in the Lamb-Dicke parameter,
the predicted cooling limit is zero. We expect next-higher order
corrections $A_+ \sim \mathcal{O}(\eta^4)$, and thus
$\langle n \rangle_{ss}^c\sim \mathcal{O}(\eta^2)$.
Also, the double-EIT configuration 
increases the cooling rate as compared to single-EIT, as the two values for
$A_-^c$ are the same in both setups, while one
has $A_+^c > 0$ for single-EIT cooling. 

In Fig.~\ref{fig-mc}(a), we show results of a numerical simulation
of the cooling dynamics for atomic parameters of a Ca$^+$ ion~\cite{ion}.
$\langle n\rangle$ is the expectation value of the vibrational quantum
number shown on a logarithmic scale against time $t$.
Curve $(i)$ shows the single-EIT setup, which is in good agreement
to both the cooling limit (dashed horizontal line) and the cooling rate 
(solid diagonal line) obtained by the single-EIT rate equation results. 
Curve $(ii)$ shows double-EIT cooling for the same
trap frequency, detuning, Lamb-Dicke parameter and cooling laser field 
intensity as in $(i)$. In this configuration, the cooling
limit is suppressed by about two powers of the Lamb-Dicke parameter $\eta$,
while the cooling rate is comparable to the single-EIT case.
In the final stage of EIT cooling close to the motional ground state,
the upper state population and thus the spontaneous emission rate is so low 
that an average over many wavefunctions is required to obtain a 
reliable prediction for the cooling limit from a quantum Monte-Carlo simulation.  
This especially holds for double-EIT cooling, where the upper state excitation
is even lower than in single-EIT cooling.
Thus we evaluated the cooling dynamics by a numerical integration of the 
full master equation including the first ten harmonic oscillator states
to avoid losses at the border of the simulated level system. The results obtained
by the numerical integration are consistent with results estimated from
full quantum Monte-Carlo simulations.
The last curve $(iii)$ shows double-EIT cooling with the same parameters as in 
$(ii)$ except for the Rabi frequencies, which are all set to 
$\Omega_i = 0.645\:\gamma_3$ ($i=1,2,3$). Thus here the cooling laser field
is not weak as compared to the driving laser fields. Whereas the cooling limit 
is similar to the single-EIT cooling case as in $(i)$, the cooling rate is 
more than one order of magnitude larger, thus allowing for a faster 
preparation of the final state.
Fig.~\ref{fig-mc}(b) shows the corresponding results for $^{199}$Hg$^+$ ions
e.g. with upper state $P_{1/2}$ ($F'=1, m_F=1$) and three lower states $S_{1/2}$
($F=1, m_F=1,0$) and $S_{1/2}$ ($F=0,m_F=0$)~\cite{apl}. 
With single-EIT cooling, the population decaying the third lower 
state has to be repumped to the
single-EIT subspace, which induces additional heating. 
The cooling dynamics for a typical setup is shown in curve~$(i)$.
Already by simply modifying the detuning of the repump field such that the
double-EIT conditions Eqs.~(\ref{c1},\ref{c3}) are fulfilled, the cooling 
limit can be lowered considerably, as shown in $(ii)$. 
Curve~$(iii)$ shows double-EIT cooling with parameters as in $(ii)$ 
except for modified Rabi frequencies of the two driving fields 
$\Omega_1, \Omega_2$, which results in a further significant lowering
of the cooling limit. (Single-EIT with Rabi frequencies as in $(iii)$,
but $\Delta_2 = 0$, yields heating of the system.)
Note that for the Hg$^+$ ion the decay rates to all
of the lower states and the additional heating due to the 
pump field have been taken into account. Also, we verified that the
improved cooling dynamics is a double-EIT effect by slightly changing the
detuning $\Delta_2$ from the optimum value $\Delta_1 - \nu$, yielding
higher cooling limits for both lower and higher detunings.

\section{Discussion and Summary}
The double-EIT scheme differs from
the conventional single-EIT scheme in that the
absorption spectrum is designed such that
both the carrier- and the blue-sideband excitations
are inhibited. This suppresses the heating processes
and the cooling limit by a factor of the order $\eta^2$.
As with the conventional single-EIT mechanism, simultaneous
cooling in three dimensions is possible if the trap frequencies
along the three axes are not too different such that 
the red-sideband absorption frequencies of all directions
are within the narrow peak in the scattering spectrum
at $\Delta_3 - \Delta_1 = \nu$. However while the double-EIT
scheme generally improves the cooling properties as compared
to the single-EIT case, a {\it simultaneous} cooling along more than
one motional axis to the ground state with efficiency as in the
one-dimensional case requires the trap frequencies to be similar, 
as otherwise the heating processes cannot be canceled exactly for all
directions at the same time. 
By coherently coupling more than three lower levels to the upper state,
the cooling scheme can also be extended to multiple-EIT cooling.
Then simultaneous EIT-cooling at different trap frequencies
 without blue-sideband heating  is possible.
Fig.~\ref{fig-absorb}(b) shows triple-EIT with four lower levels,
where $n_1$ ($n_2$) is the motional quantum number for trap frequency
$\nu_1$ ($\nu_2$). 
Alternatively, multiple-EIT can be used to cancel further heating processes.
However at higher order in $\eta$, also the heating via scattering of the cooling 
fields needs to be taken into account.

It should be noted that in some systems double-EIT cooling does not
require more effort than single-EIT cooling. Typical examples are systems
 where the upper state decays to more than two lower states.
Instead of repumping the system arbitrarily as required 
for single-EIT cooling, a suitable choice for the
parameters of the pump laser field leads to double-EIT cooling and thus 
to improved cooling performance. Therefore here the disadvantage of 
additional  decay channels becomes an advantage. 
Other possible systems for double-EIT cooling are 
e.g. Ne~\cite{neon}, Rb or Cs. Depending on the choice of the upper 
state, the latter two systems also allow for multiple-EIT cooling with more
than three lower states.

In summary, we have presented a laser cooling scheme
which allows to efficiently reach the motional ground state of
a trapped atom or ion. 
With the help of two coupling laser fields,
the scattering spectrum for the cooling laser
field is designed such that both carrier- and blue-sideband
excitations are inhibited, thus eliminating all heating
processes in leading order of an expansion in the Lamb-Dicke 
parameter $\eta$. As compared to the conventional single-EIT scheme,
this double-EIT scheme allows to lower the cooling limit
by a factor of order $\eta^2$.

\begin{acknowledgments}
Financial support is acknowledged by the 
German National Academic Foundation for JE.
\end{acknowledgments}


\begin{thebibliography}{30}
\bibitem{erste} W. Neunhauser, M. Hohenstatt, P. Toschek and
H. Dehmelt, Phys. Rev. Lett. {\bf 41}, 233 (1978);
D. J. Wineland, R. E. Drullinger, and F. L. Walls,
Phys. Rev. Lett. {\bf 40}, 1639 (1978).
\bibitem{jumps} W. Nagourney, J. Sandberg, and H. Dehmelt,
Phys. Rev. Lett. {\bf 56}, 2797 (1986);
Th. Sauter, W. Neuhauser, R. Blatt, and P. E. Toschek,
Phys. Rev. Lett. {\bf 57}, 1696 (1986);
J. C. Bergquist, R. G. Hulet, W. M. Itano, and D. J. Wineland,
Phys. Rev. Lett. {\bf 57}, 1699 (1986).
\bibitem{ground-state} F. Diedrich, J. C. Bergquist, W. M. Itano, 
and D. J. Wineland, Phys. Rev. Lett. {\bf 62}, 403 (1989).
\bibitem{spectroscopy} S. Diddams et al., Science {\bf 293}, 825 (2000).
\bibitem{forbidden} Ch. Roos et al., Phys. Rev. Lett. {\bf 83},
4713 (1999).
\bibitem{raman-cooling} C. Monroe et al., Phys. Rev. Lett. {\bf 75},
4011 (1995).
\bibitem{eit} G. Morigi, J. Eschner and C. H. Keitel, Phys. Rev. Lett.
{\bf 85}, 4458 (2000).
\bibitem{eit2} G. Morigi, Phys. Rev. A {\bf 67}, 033402 (2003).
\bibitem{eit-exp} C. F. Roos, D. Leibfried {\it et al.}, Phys. Rev. Lett. {\bf 85}, 5547 (2000).
\bibitem{hot} A. Sorensen and K. Molmer, Phys. Rev. Lett. {\bf 82}, 1971 (1999); D. Jonathan and M. B. Plenio, Phys. Rev. Lett. {\bf 87}, 127901 (2001);  
 J. J. Garcia-Ripoll, P. Zoller, J. I. Cirac,
Phys. Rev. Lett. {\bf 91}, 157901 (2003); S.-B. Zheng, Phys. Rev. Lett.
{\bf 90}, 217901 (2003).
\bibitem{n-lower-state}E. Paspalakis and P.L. Knight, Phys. Rev. A {\bf 66}, 
015802 (2002).
\bibitem{rmp}D. Leibfried, R. Blatt, C. Monroe, D. Wineland, 
Rev. Mod. Phys. {\bf 75}, 281 (2003).
\bibitem{review2} J. Eschner, G. Morigi, F. Schmidt-Kahler and R. Blatt,
J. Opt. Soc. Am. B {\bf 20}, 1003 (2003).
\bibitem{apl} F. Schmidt-Kaler {\it et al.}, Appl. Phys. B {\bf 73},
807 (2001).
\bibitem{ion}H.C. N\"agerl {\it et al.}, Appl. Phys. B {\bf 66}, 603 (1998).
\bibitem{neon}H. Theuer {\it et al.}, Optics Express {\bf 4}, 77 (1999).
\end{thebibliography}
\end{document}